\documentclass[cits]{PoS}
\pdfoutput=1
\usepackage{amsmath}
\usepackage[utf8]{inputenc}
\def\gev{\,\text{Ge\hspace{-0.1em}V}}

\makeatletter
     {\bgroup\raggedright\small\section*{\refname
        \@mkboth{\MakeUppercase\refname}{\MakeUppercase\refname}}%
      \list{\name{bib\@arabic\c@enumiv}
	    \@biblabel{\@arabic\c@enumiv}}%
           {\settowidth\labelwidth{\@biblabel{#1}}%
            \setlength\itemsep{0pt}
            \setlength\parskip{0pt}
            \setlength\parsep{0pt}
            \setlength\partopsep{0pt}
            \leftmargin\labelwidth
            \advance\leftmargin\labelsep
            \@openbib@code
            \usecounter{enumiv}%
            \let\p@enumiv\@empty
            \renewcommand\theenumiv{\@arabic\c@enumiv}}%
      \sloppy\clubpenalty4000\widowpenalty4000%
      \sfcode`\.\@m}
     {\def\@noitemerr
       {\@latex@warning{Empty `thebibliography' environment}}%
      \endlist\egroup}
\makeatother

\title{Semileptonic $B\to\pi\ell\nu$, $B\to D\ell\nu$, $B_s\to K\ell\nu$,  and $B_s\to D_s\ell\nu$ decays}

\ShortTitle{Semileptonic $B\to\pi\ell\nu$, $B\to D\ell\nu$, $B_s\to K\ell\nu$,  and $B_s\to D_s\ell\nu$ decays}

\author{{Jonathan Flynn}\\
        Physics and Astronomy, University of Southampton, Southampton SO17 1BJ, UK}

\author{\speaker{Ryan Hill}\\
        Physics and Astronomy, University of Southampton, Southampton SO17 1BJ, UK\\
        DISCnet Centre for Doctoral Training, University of Southampton, Southampton SO17 1BJ, UK\\
        E-mail: \email{R.C.Hill@soton.ac.uk}}
        
        \author{{Andreas J\"{u}ttner}\\
        Physics and Astronomy, University of Southampton, Southampton SO17 1BJ, UK\\
        STAG  Research  Center  and  Mathematical  Sciences, University  of  Southampton,\\
        Southampton  SO17  1BJ,  UK}
        
        \author{Amarjit Soni\\
        Physics Department, Brookhaven National Laboratory, Upton, NY 11973, USA}
        
        \author{{Justus Tobias Tsang}\\
        Higgs Centre for Theoretical Physics, The University of Edinburgh, EH9 3FD, UK\\
        CP3-Origins and IMADA, University of Southern Denmark, Campusvej 55, 5230 Odense M, Denmark}
        
        \author{Oliver Witzel$^*$\\
        Department of Physics, University of Colorado Boulder, Boulder, CO 80303, USA\\
        E-mail: \email{Oliver.Witzel@colorado.edu}}

        \abstract{We present updates for our nonperturbative lattice QCD
          calculations to determine semileptonic form factors for exclusive
          $B\to \pi\ell\nu$, $B\to D \ell\nu$, $B_s\to K\ell \nu$, and $B_s\to
          D_s\ell\nu$ decays. Our calculation is based on RBC-UKQCD's set of
          $2+1$-dynamical-flavor gauge field ensembles. In the valence sector
          we use domain wall fermions for up/down, strange and charm quarks,
          whereas bottom quarks are simulated with the relativistic heavy quark
          action. The continuum limit is based on three lattice spacings.
          Using kinematical $z$ expansions we aim to obtain form factors over
          the full $q^2$ range.  These form factors are
          the basis for predicting ratios addressing lepton flavor universality
          or, when combined with experimental results, to obtain CKM matrix
          elements $|V_{ub}|$ and $|V_{cb}|$.}

\FullConference{The 37th Annual International Symposium on Lattice Field Theory -- LATTICE2019\\
16--22 June 2019\\
Wuhan, China}

\begin{document}

\section{Introduction}
The Standard Model (SM) of elementary particle physics has been tested
in numerous processes over recent decades and the overall agreement
between theoretical predictions and experimental observations is
astonishing. The SM successfully describes three of the four
fundamental forces in nature: electromagnetic, weak, and strong, with
only gravitation not included. Although successful, the SM is an
effective theory and corrections due to `new physics' are expected to
occur at higher energies. In order to observe such effects, one
possibility is to perform precision tests of SM-allowed processes,
which requires both experimental and theoretical
effort~\cite{Tanabashi:2018oca}. Particularly interesting are weak
quark-flavor-changing transitions.

Quark flavor mixing is typically described by the
Cabibbo-Kobayashi-Maskawa (CKM) matrix which is a unitary $3\times 3$
matrix in the SM. Testing the unitarity of the CKM matrix remains a
major focus of the experimental and theoretical research program.
Decays involving bottom (or $b$) quarks are most promising because,
compared to the light sector, theoretical predictions are currently less precise
and the much larger mass of the $b$ quark allows many decay channels.
Experimentally, interactions of $b$ quarks can be studied by looking
for $B$-mesons and their decays. Two current experiments, Belle
II~\cite{Kou:2018nap} and LHCb~\cite{Bediaga:2018lhg}, are dedicated
to $B$-physics but ATLAS and CMS also study such processes. At present
the $b$ sector of the SM exhibits various tantalizing tensions between
experimental measurements and theoretical predictions. Most noteworthy
are the signs of lepton flavor universality violation observed in
semileptonic decays. $B \to D^{(*)} \ell \nu$ decays probe the
transition of a $b$ quark to a $c$ quark, and forming the ratio of
processes with a heavy $\tau \nu_\tau$ lepton pair in the final state
to those containing the much lighter $\mu \nu_\mu$ (or $e\nu_e$),
defines the $R$-ratio
\begin{align}
  R_{D^{(*)}}^{\tau/\mu}\equiv \frac{BF(B\to D^{(*)}\tau\nu_\tau)}{BF(B\to D^{(*)} \mu \nu_\mu)}.
\end{align}
Experimentally this ratio has been determined by BaBar, Belle and LHCb
\cite{Lees:2012xj,Lees:2013uzd,Huschle:2015rga,Aaij:2015yra,Sato:2016svk,Hirose:2016wfn,Hirose:2017dxl,Aaij:2017uff,Aaij:2017deq,Abdesselam:2019dgh}
and the SM predictions are based on
\cite{Bigi:2016mdz,Bernlochner:2017jka,Bigi:2017jbd,Jaiswal:2017rve}.
The average experimental and theoretical values exhibit at present a
discrepancy of about $3\sigma$ \cite{HFLAV:RdRds-2019}. Since many
uncertainties cancel, this ratio is a theoretically clean quantity and
hence a strong test of the SM. Furthermore, these semileptonic decays
allow extraction of the CKM matrix element $|V_{cb}|$ by combining
experimental and theoretical results. The determination based on these
exclusive channels exhibits a $2$--$3\sigma$ tension with that from
inclusive $B\to X_c\ell\nu$ where $X_c$ denotes any final state
containing a charm
quark~\cite{Charles:2004jd,Bona:2005vz,Antonelli:2009ws,Tanabashi:2018oca}.
A similar tension is observed for the CKM matrix element $|V_{ub}|$
where the exclusive $B\to \pi \ell \nu$ and the inclusive $B\to X_u
\ell \nu$ channels are typically considered, with $X_u$ being a
charmless final state containing an up-quark.

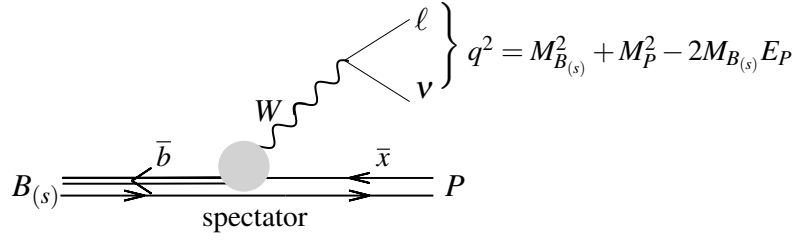
\begin{figure}[tb]
  \centering
  \begin{picture}(100,30)
    \put(6,2){\includegraphics[width=50mm]{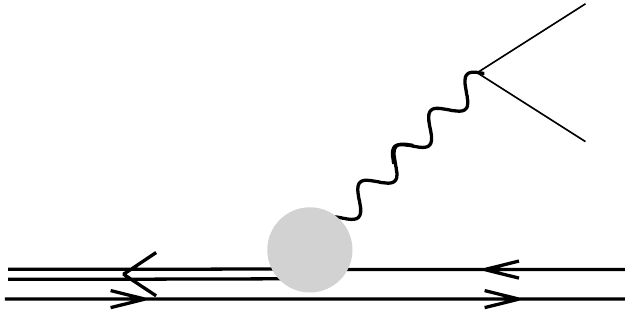}}
    \put(0,3){\large{$B_{(s)}$}} \put(57,3){\large{$P$}}
    \put(32,13){\large{$W$}} 
    \put(53,25){\large{$\ell$}} \put(53,16){\large{$\nu$}}
    \put(56,22){$\Bigg\}$}\put(60,21){$q^2= M_{B_{(s)}}^2 + M_P^2 - 2 M_{B_{(s)}} E_P$}
    \put(25,-1){spectator}
    \put(48,7){$\overline{x}$}
    \put(19,7){$\overline{b}$}
  \end{picture}
  \caption{Sketch of tree-level weak semileptonic $B_{(s)}$ decays
    mediated by a charged $W^\pm$ boson in a set-up with the $B_{(s)}$
    meson at rest. $P$ denotes a pseudoscalar final state ($\pi$, $K$,
    $D$, or $D_s$), the spectator is a light up/down or a strange
    quark, and $x$ the up or charm daughter quark. }
  \label{fig.treelevel}
\end{figure}

These investigations are based on branching fractions (BF) which are experimentally measured and conventionally parameterized by
\begin{multline}
  \frac{d\Gamma(B_{(s)}\to P\ell\nu)}{dq^2} = \frac{\eta_{EW} G_F^2 |V_{xb}|^2}{24\pi^3}\frac{(q^2-m_\ell^2)^2\sqrt{E^2_P - M^2_P}}{q^4 M_{B_{(s)}}^2} \\
  \times \Bigg[ \bigg(\!1 + \frac{m^2_\ell}{2q^2}\! \bigg)M^2_{B_{(s)}}(E_P^2 - M_P^2) |f_+(q^2)|^2 + 
    \frac{3m^2_\ell}{8q^2}(M_{B_{(s)}}^2 - M_P^2)^2 |f_0(q^2)|^2  \Bigg]\,.
  \label{Eq.BF}
\end{multline}
In Eq.~(\ref{Eq.BF}) we focus solely on semileptonic
$B_{(s)}$ decays mediated by a charged current with a pseudoscalar particle
($P$) in the final state as sketched in Fig.~\ref{fig.treelevel}. $M_P$ is the
mass of this particle, $E_P$ its energy, and $|V_{xb}|$ is the CKM matrix of
interest with $x=u,c$. The four-momentum transferred to the leptonic products of
type $\ell=e,\,\mu,\,\tau$ is denoted by $q^\mu$ and $m_\ell$ is the mass of the
lepton. On the theoretical side, the determination faces the challenge that
nonperturbative interactions due to the strong force contribute, parametrized by
the form factors $f_+$ and $f_0$, whereas $G_F$ is the perturbatively-computed
Fermi constant. Lattice quantum chromodynamics (QCD) provides a state-of-the-art
nonperturbative framework for determining the form factors. An overview of
existing results based on lattice QCD as well as averages obtained from them can
be found in~\cite{Aoki:2019cca}.  Here we focus on four
semileptonic $B_{(s)}$ decay channels which have been calculated by

\smallskip
  \begin{tabular}{l@{}ll}
  & $B   \to \pi \ell\nu$: & HPQCD \cite{Dalgic:2006dt}, Fermilab/MILC \cite{Bailey:2008wp,Lattice:2015tia}, RBC/UKQCD\cite{Flynn:2015mha}\\
  & $B   \to D   \ell\nu$: & Fermilab/MILC \cite{Bailey:2012jg,Lattice:2015rga}, HPQCD \cite{Na:2015kha}\\
  & $B_s \to K   \ell\nu$: & HPQCD \cite{Bouchard:2014ypa},  RBC/UKQCD\cite{Flynn:2015mha}, Alpha \cite{Bahr:2016ayy}, Fermilab/MILC \cite{Bazavov:2019aom}\\
  & $B_s \to D_s \ell\nu$: & Fermilab/MILC \cite{Bailey:2012rr,Bazavov:2019aom}, HPQCD \cite{Monahan:2017uby,McLean:2019qcx} \\
  \end{tabular}
  \smallskip
  
\noindent In addition several groups report on work in progress \cite{Hashimoto:2019pgh,Colquhoun:2018kwj,Lubicz:2018scy,Kaneko:2018mcr,Colquhoun:2019tyq}; see also work referred in \cite{Lytle:2019plenary}.

As indicated in Fig.~\ref{fig.treelevel}, all four processes change
the flavor of the $b$ quark in the initial state by emission of a
charged $W^\pm$ boson. The spectator quark is an up or down quark for $B$
decays and a strange quark for $B_s$ decays. The daughter quark is
either an up or charm quark to allow $\pi$, $D$ or $K$, $D_s$ final
states, respectively.

We present these four channels together because in our set-up only the
bottom quark is simulated by the relativistic heavy quark (RHQ) action
\cite{ElKhadra:1996mp,Christ:2006us} whereas up/down, strange, and
charm quarks are simulated using domain wall fermions (DWF)
\cite{Kaplan:1992bt,
  Shamir:1993zy,Furman:1994ky,Blum:1996jf,Blum:1997mz,Brower:2012vk}.
We use the RBC-UKQCD $2+1$-flavor domain wall fermion and Iwasaki
gauge action ensembles
\cite{Allton:2008pn,Aoki:2010dy,Blum:2014tka,Boyle:2017jwu}, and list
their key properties in Tab.~\ref{Tab.ens}. The F1 ensemble is a new
addition compared to our previous
publications~\cite{Christ:2014uea,Flynn:2015mha,Flynn:2015xna}. It
provides a third, finer lattice spacing as well as a data point closer
to the physical pion mass. This further constrains the extrapolations
to the continuum limit and to the physical light quark mass. In the
valence sector we use the unitary light quark mass for $u/d$ quarks
and choose a close-to-physical value for the strange quark mass. Charm
quarks are simulated using a domain wall action optimized for heavy
flavors~\cite{Cho:2015ffa,Boyle:2016imm}. For the coarse ensembles
(C1, C2), three `lighter-than-charm' masses are simulated and we
subsequently perform a benign extrapolation, whereas for the medium
fine (M1, M2, M3) and fine (F1) ensembles we bracket the physical
charm quark mass and interpolate. As mentioned above, bottom quarks
are simulated with the effective RHQ action, with nonperturbatively
tuned parameters~\cite{Aoki:2012xaa} which have been updated compared
to our previous work to reflect improved determinations of the lattice
spacing.
  
\begin{table}[t]
\begin{center}
\begin{tabular}{ccccccc}
\hline
& $L^3 \times T$ & $a^{-1}$ / GeV & $am_l$ & $M_\pi$ / MeV & \# Configurations & \# Time Sources\\
\hline\hline
C1& $24^3 \times 64$&1.784 & 0.005& 338& 1636& 1\\
C2& $24^3 \times 64$&1.784 & 0.010& 434& 1419& 1\\
M1& $32^3 \times 64$&2.383 & 0.004& 301& 628& 2\\
M2& $32^3 \times 64$&2.383 & 0.006& 362& 889& 2\\
M3& $32^3 \times 64$&2.383 & 0.008& 411& 544& 2\\
F1& $48^3 \times 96$&2.774 & 0.002144& 234& 98& 24\\
\hline
\end{tabular}
\end{center}
\caption{The RBC-UKQCD 2+1 domain-wall fermion ensembles used in this work \cite{Allton:2008pn,Aoki:2010dy, Blum:2014tka, Boyle:2017jwu}. The F1 ensemble is a new element of the RBC/UKQCD $b$-physics project analysis and is a key difference between this work and the prior analysis. Presently the properties of the F1 ensemble are re-evaluated and may change slightly.}
\label{Tab.ens}
\end{table}

The form factors introduced in Eq.~(\ref{Eq.BF}) can be determined by
computing the hadronic matrix elements $\langle P | {\cal V}^\mu |
B_{(s)} \rangle$, parameterized as
\begin{align}
\langle P(p_P) | {\cal V}^\mu |B_{(s)}(p_{B_{(s)}})\rangle = f_+(q^2)\left( p^\mu_{B_{(s)}} + p^\mu_P - \frac{M_{B_{(s)}}^2 - M_P^2}{q^2}q^\mu \right) + f_0(q^2)\frac{M_{B_{(s)}}^2 - M_P^2}{q^2}q^\mu\,,
\end{align}
with $p_{B_{(s)}}$ and $p_P$ the momenta of the $B_{(s)}$ and $P$
mesons respectively, and ${\cal V}^\mu = \bar{x}\gamma^\mu b$. In our
calculation we place the initial $B_{(s)}$ meson at rest and use an
alternative parameterization of the matrix elements in terms of the
perpendicular and parallel components 
\begin{align}
\langle P | {\cal V}^\mu |B_{(s)}\rangle = \sqrt{2M_{B_{(s)}}}\left[ v^\mu f_\parallel(E_P) + p_\perp^\mu f_\perp(E_P)\right]\,,
\end{align}
with $p_\perp^\mu\equiv p^\mu_P-(p_P\cdot v)v^\mu$. Hence we determine the form factors
\begin{align}
  f_\parallel = \frac{\langle P | {\cal V}^0 |B_{(s)}\rangle}{\sqrt{2M_{B_{(s)}}}}, \qquad\qquad
  f_\perp = \frac{\langle \pi | {\cal V}^i |B_{(s)}\rangle}{\sqrt{2M_{B_{(s)}}}} \frac{1}{p^i_{P}}
  \label{Eq.f_lat}  
\end{align}
and obtain the phenomenological form factors $f_+$ and $f_0$ from
\begin{align}
f_+(q^2) &= \frac{1}{\sqrt{2M_{B_{(s)}}}} \left[f_\parallel(q^2) + (M_{B_{(s)}} - E_P)f_\perp(q^2) \right],\\
f_0(q^2) &= \frac{\sqrt{2M_{B_{(s)}}}}{M^2_{B_{(s)}} - M^2_P} \left[(M_{B_{(s)}} - E_P)f_\parallel(q^2) + (E^2_P - M^2_P)f_\perp(q^2) \right].
\end{align}
For the lattice determination we evaluate Eq.~(\ref{Eq.f_lat}) by computing
three-point and two-point correlation functions on the lattice.
The operators entering the three-point functions include tree-level plus
1-loop $O(a)$-improved terms with perturbatively calculated improvement
coefficients. The hadronic matrix elements are obtained by performing
correlated fits to ratios of these three-point and two-point functions
using the bootstrap resampling technique. In order to test
for systematic effects due to excited state contributions, we fit to a
simple plateau assuming only ground state contributions, but also with
additional terms to describe one added excited state for the initial
as well as the final state. The matrix elements containing the vector
current ${\cal V}_\mu$ are finally renormalized to obtain continuum expressions
\begin{align}
\langle P|{\cal V}_\mu|B_{(s)}\rangle = Z_\mu^{bx} \langle P|V_\mu|B_{(s)} \rangle \,.
\end{align}
The heavy-light renormalization factor $Z_\mu^{bx}$ is obtained using a
`mostly nonperturbative' ansatz \cite{ElKhadra:2001rv}
\begin{align}
Z_{V_\mu}^{bx}= \rho_{V_\mu}^{bx} \sqrt{Z_\mu^{bb} Z_\mu^{xx}},
\end{align}
where the flavor conserving factors $Z_\mu^{bb}$ and $Z_\mu^{xx}$ are
computed nonperturbatively and the remaining factor
$\rho_{V_\mu}^{bx}$ is determined at $1$-loop in lattice perturbation
theory.

In the following Section we will report on the status of the four
decay channels. We start with our more preliminary determinations of
form factors for $B\to\pi\ell\nu$ and $B\to D \ell \nu$ and continue
with the more advanced results for $B_s\to K\ell\nu$ and $B_s\to
D_s\ell\nu$. In Section \ref{Sec.zfits} we present details of the
kinematical $z$-expansion followed by a brief summary.
The determinations of $B\to\pi\ell\nu$ and $B_s\to K\ell\nu$ form
factors update our results from 2015~\cite{Flynn:2015mha}.

\section{Semileptonic form factors}
\subsection{$B\to\pi\ell\nu$} \label{Sec.Bpi}

\begin{figure}[tb] 
\includegraphics[width=0.48\textwidth]{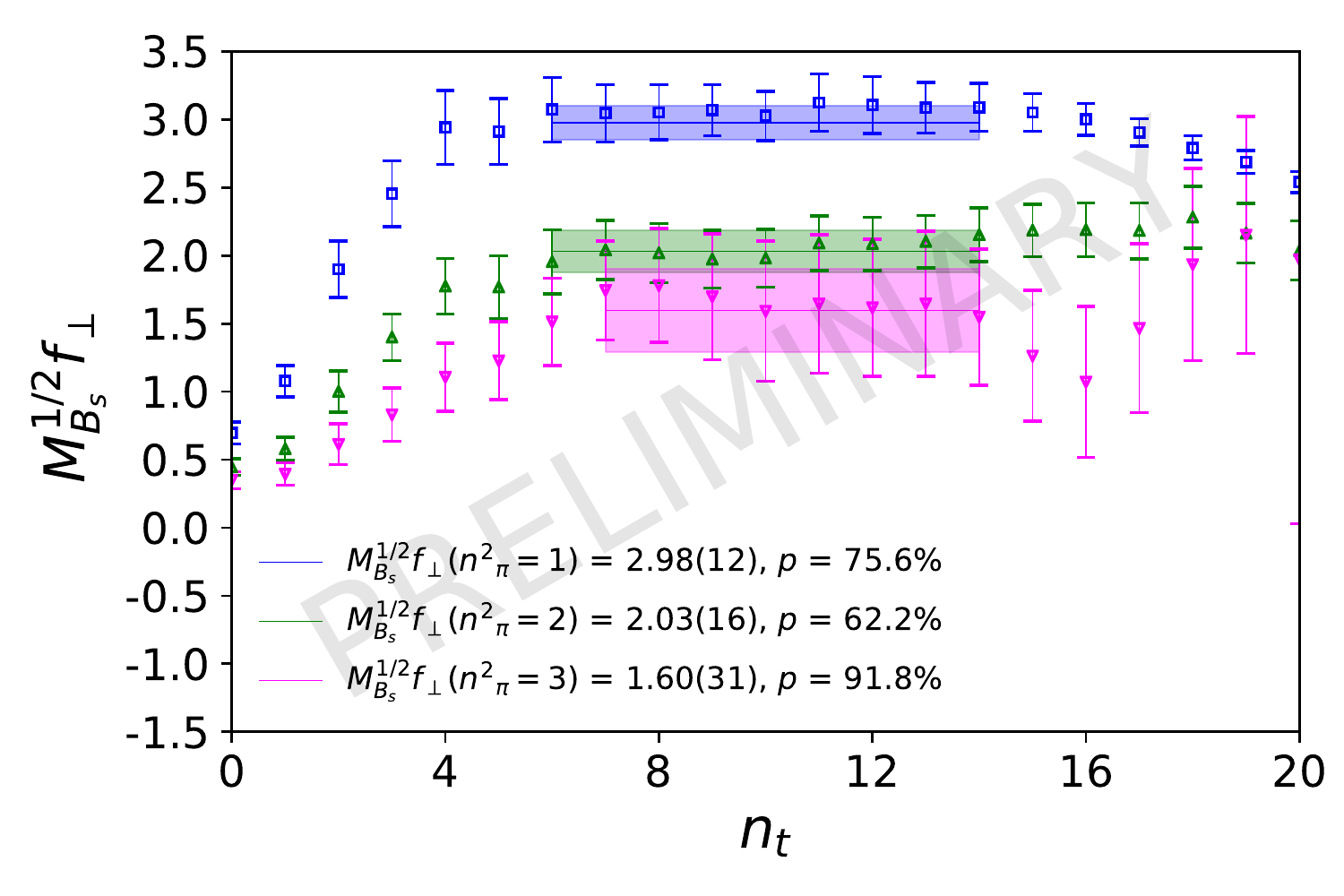}
\includegraphics[width=0.48\textwidth]{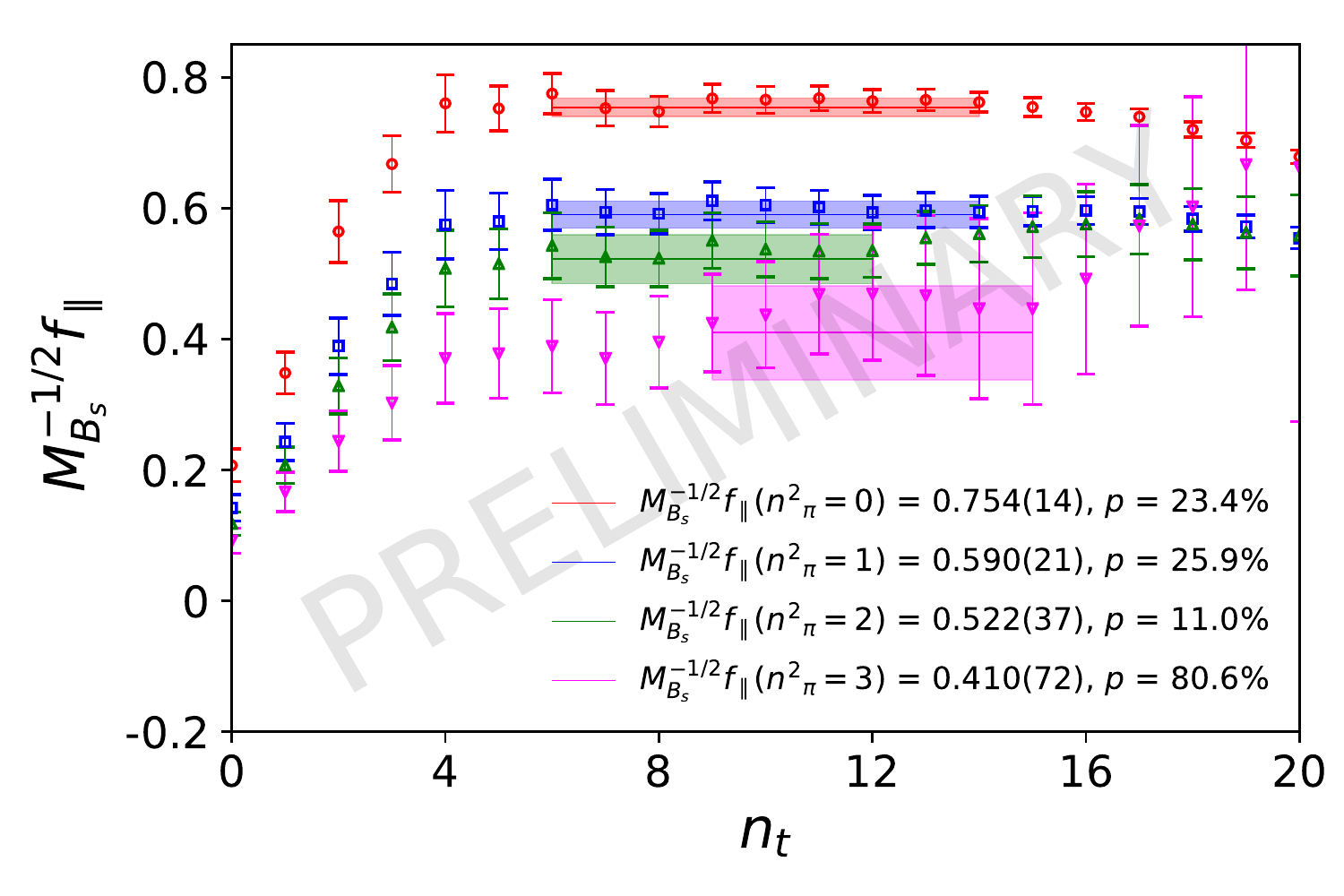}
\caption{Plateau fits to ratios of three-point and two-point functions used to determine $f_\perp$ and $f_\parallel$ on the M1 ensemble for $B\to\pi\ell\nu$ semileptonic decays.}
\label{fig.BpiFormFactorsM1}
\end{figure}
\begin{figure}[tb]
\includegraphics[width=0.48\textwidth]{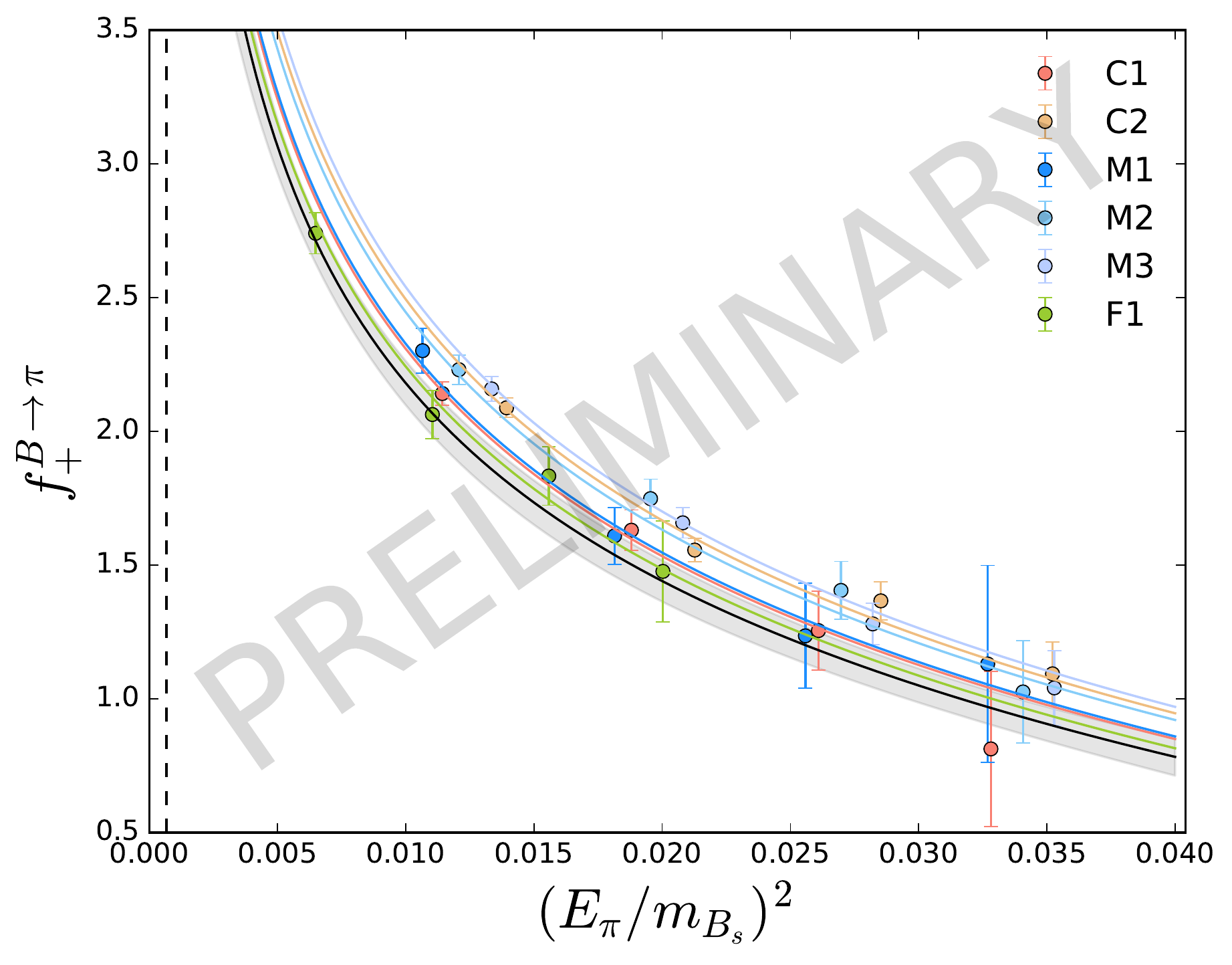}
\includegraphics[width=0.48\textwidth]{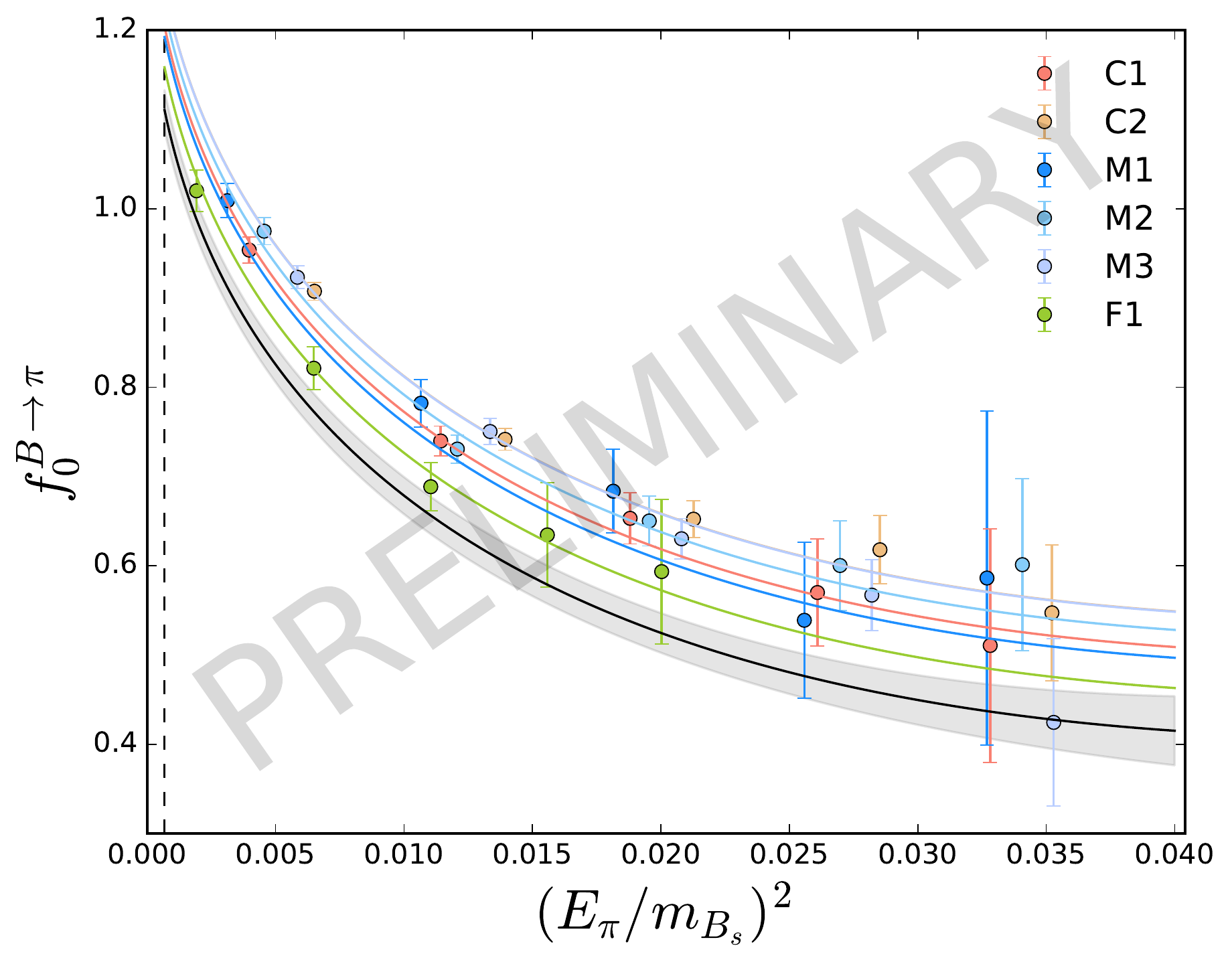}
\caption{$B\to\pi\ell\nu$ chiral-continuum fit to the $f_+$ and $f_0$ data using heavy meson chiral perturbation theory. The determination of the form factors at discretized momenta for our six ensembles are shown by the colored symbols and the chiral-continuum limit is denoted by the black line with gray error band. Only statistical errors are shown in the plots.}
\label{fig.BpiFormFactors}
\end{figure}
Following the steps outlined above, we first determine $f_\parallel$
and $f_\perp$ on all six of our ensembles. Figure
\ref{fig.BpiFormFactorsM1} shows an example of fits to extract the plateau values. Having determined and renormalized the form factors from the lattice data, we obtain the chiral-continuum limit by performing correlated global fits to all data points. The fit function is derived from SU(2) heavy meson chiral perturbation theory in the hard-pion limit \cite{Becirevic:2002sc, Bijnens:2010ws}
\begin{align}
  f^{B\to\pi}(M_\pi, E_\pi, a) = \frac{1}{E_\pi + \Delta} c_0 \left[ 1 + \frac{\delta f^{B\to\pi}}{(4\pi f_\pi)^2} + c_1\frac{M_\pi^2}{\Lambda^2} + c_2\frac{E_\pi}{\Lambda} + c_3\frac{E_\pi^2}{\Lambda^2} + c_4 (a\Lambda)^2\right].
  \label{Eq.HMXPT}
\end{align}
In the equation above, $\Delta = M_{B^*} - M_B$ and $B^*$ is a $\bar{b}u$ flavor
state with quantum numbers $J^P=1^-$ for $f_+$, or $J^P=0^+$ for $f_0$. Since for $B\to\pi\ell\nu$ decays only the vector $B^*$ meson lies below the production threshold, we include the pole factor $1/(E_\pi+\Delta)$ only for $f_+$ but not for $f_0$. The non-analytic term is
\begin{align}
\delta f^{B\to\pi} = -\frac{3}{4}(3g_b^2 + 1)M_\pi^2 \log\left(\frac{M_\pi^2}{\Lambda^2}\right)\,,
\end{align}
where $f_\pi$ is the pion decay constant, $\Lambda$ a reference scale of $O(1\gev)$,
and $g_b$ is the $B^*B\pi$ coupling constant determined in Ref.~\cite{Flynn:2015xna}.  The
final chiral-continuum limit for $f_+$ and $f_0$ is shown by the black line with
gray error band in the plots in Fig.~\ref{fig.BpiFormFactors} whereas the
colored data points are our renormalized lattice data and the colored lines show
the fit results at finite lattice spacing and unphysical $u$/$d$ quark
masses. The plots in Fig.~\ref{fig.BpiFormFactors} cover the energy range we can
directly simulate on our set of ensembles. Exploring more energetic pions is
challenging because the signal-to-noise ratio deteriorates as the pion momentum
increases.

We therefore construct a set of synthetic data points from the
continuum description of the form factors and estimate all systematic
uncertainties for these points. A complete error budget can then be
fed into a kinematical $z$-expansion to obtain form factors covering
the full range of physically allowed momentum transfer $q^2$. Details
of the $z$-expansion are described in Sec.~\ref{Sec.zfits}.

\subsection{$B\to D\ell\nu$}\label{Sec.BD}

\begin{figure}[tb]
  \includegraphics[width=0.48\textwidth]{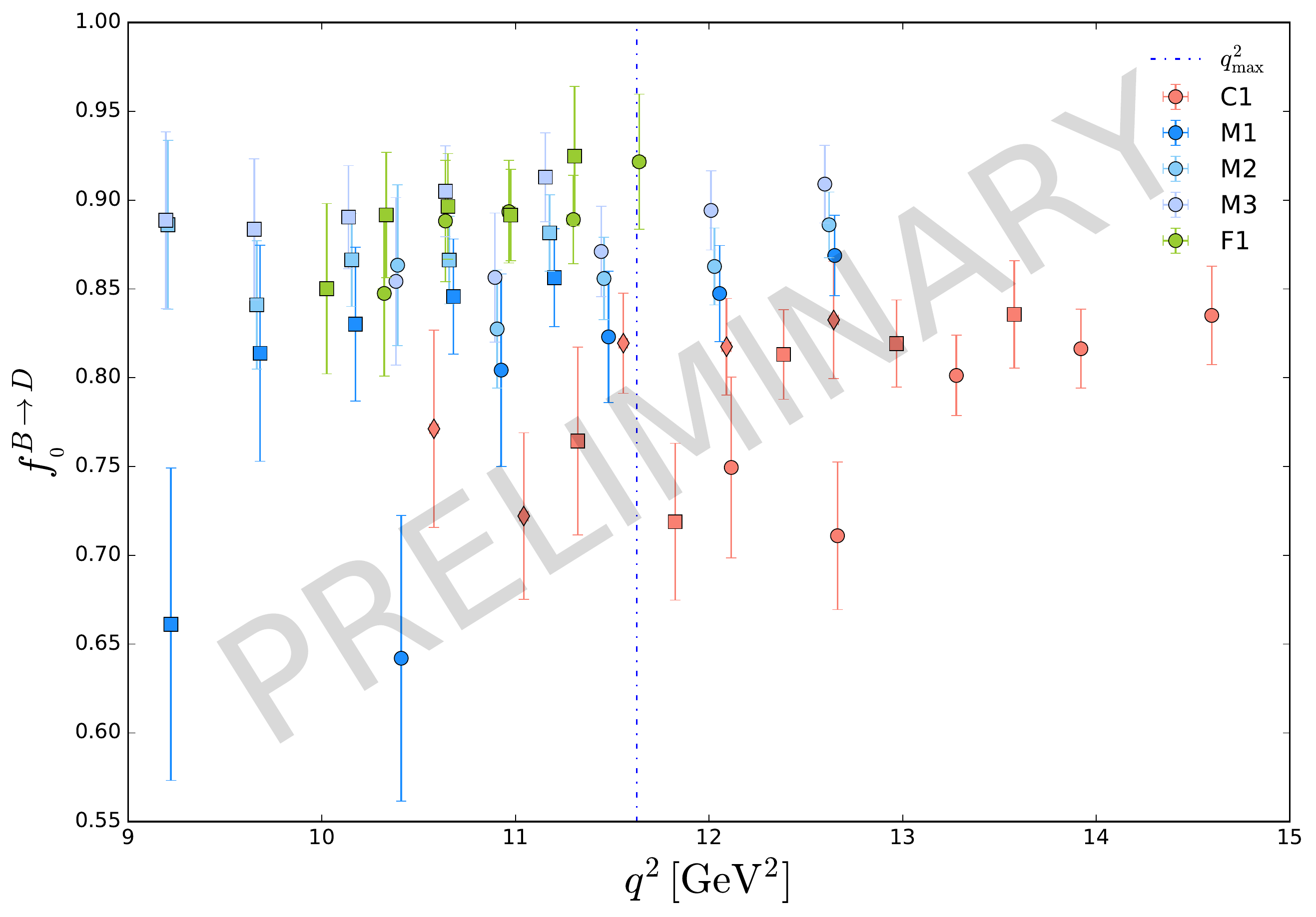}
  \includegraphics[width=0.48\textwidth]{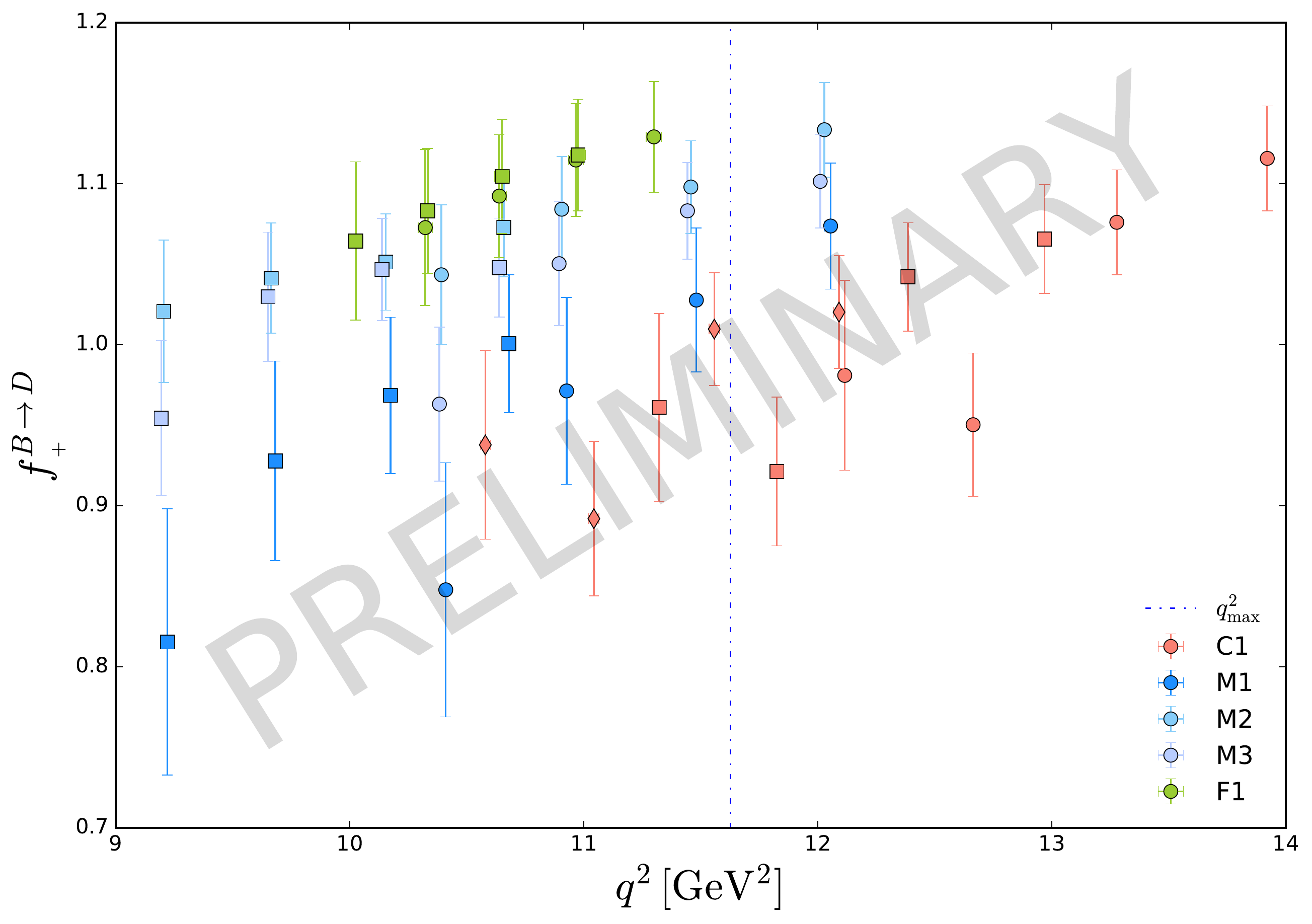}
\caption{Form factors $f_+$ and $f_0$ for $B\to D\ell\nu$ semileptonic decays. Since we either extrapolate three `lighter-than-charm' quark masses or interpolate bracketing charm quark masses, several sets of data points are shown for each ensemble.}
\label{fig.BDFormFactors}
\end{figure}

The determination of $B\to D \ell\nu$ form factors requires us to
simulate charm quarks. We do so using DWF optimized for heavy
quarks~\cite{Boyle:2016imm}. On the coarse ensembles this formulation
does not allow us to reach physical charm quark masses and we
therefore simulate three values lighter than the charm quark mass. On
the medium fine and fine ensembles, however, we do reach the physical
charm quark mass and bracket its value. In
Fig.~\ref{fig.BDFormFactors} we show the renormalized form factors
$f_+$ and $f_0$ for five of our six ensembles and the set of
three or two charm quark masses. 

As for $B\to \pi\ell \nu$ the next step is to perform a
chiral-continuum extrapolation to the continuum and to physical light
quark masses by performing a fit to all data points. In addition, we
need to include an extra(inter)polation to the physical charm quark
mass. We are currently investigating different fit ans\"atze and also
compare a global fit performing all extrapolations at once compared to
a two step fit separating the charm extra(inter)polation from the
chiral-continuum extrapolation.

\subsection{$B_s\to K \ell \nu$}\label{Sec.BsK}

\begin{figure}[tb]
  \includegraphics[width=0.48\textwidth]{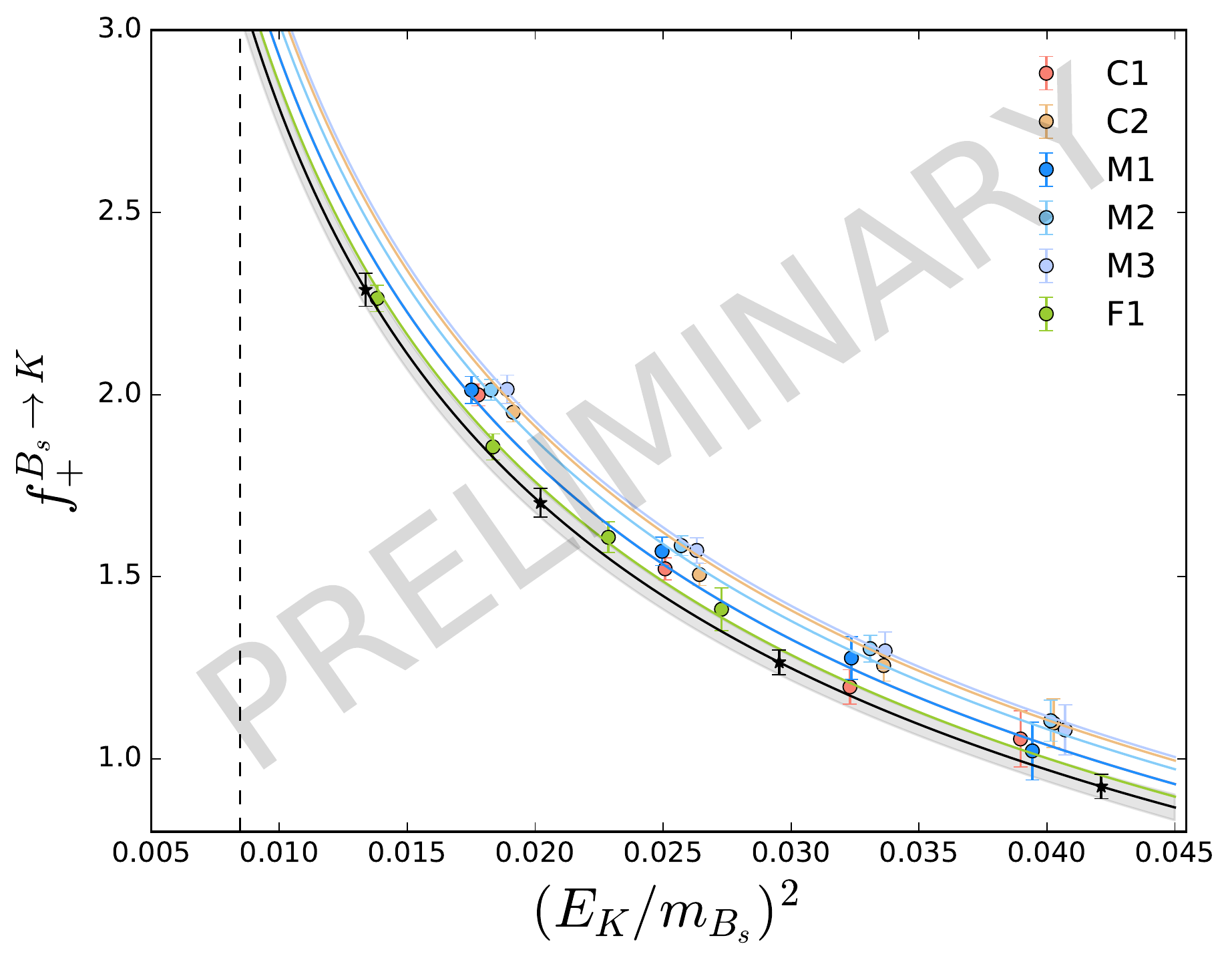}
  \includegraphics[width=0.48\textwidth]{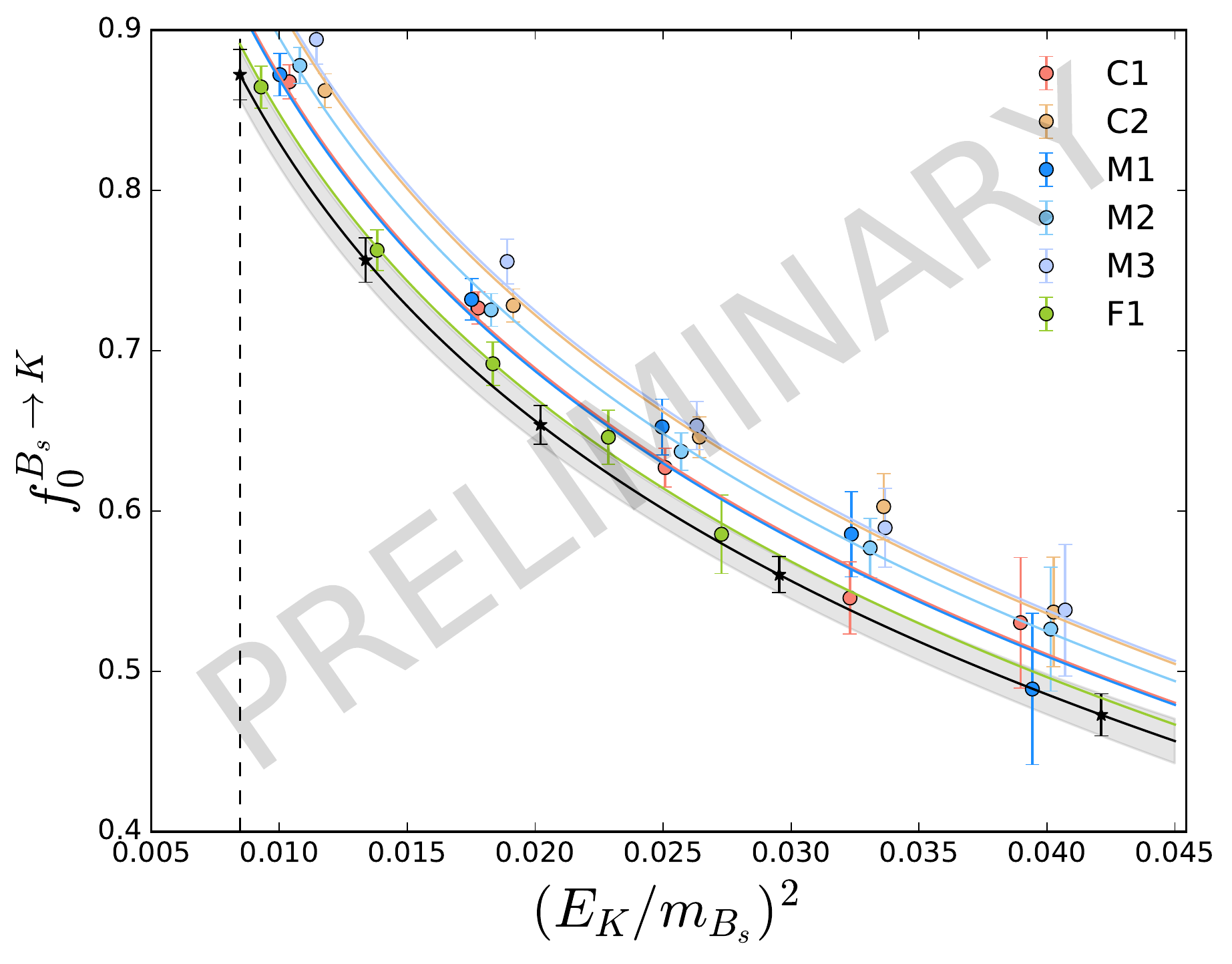}
  \caption{Chiral-continuum extrapolation for $B_s\to K \ell\nu$ decays using SU(2) hard-kaon heavy meson $\chi$PT. Colored symbols show the form factors determined on our six ensembles and the black line with gray error band the result of the chiral-continuum extrapolation. Only statistical uncertainties are shown.}
  \label{fig.BsK_ChiPT}
\end{figure}

\begin{figure}[tb]
  \includegraphics[width=0.48\textwidth]{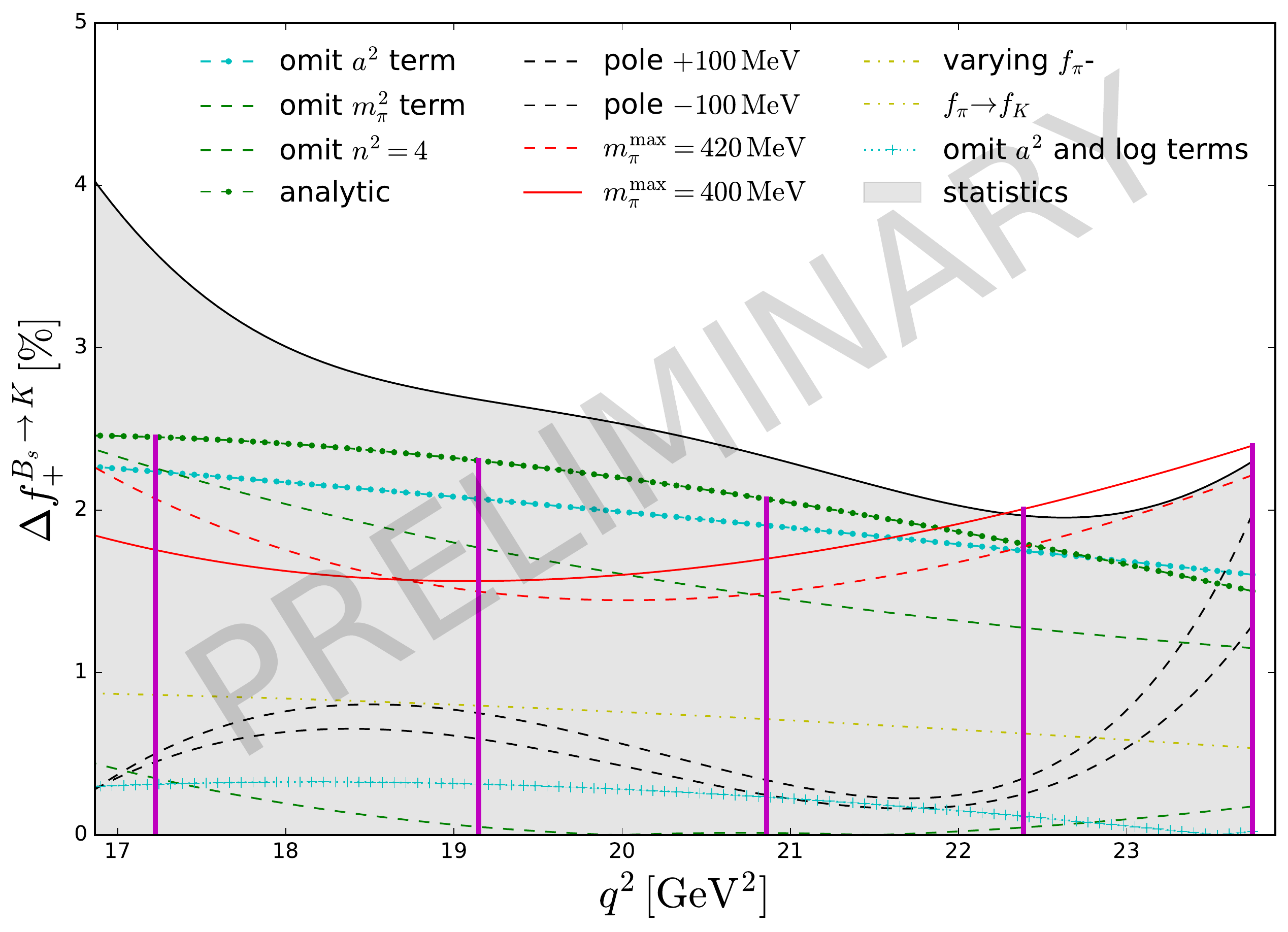}
  \includegraphics[width=0.48\textwidth]{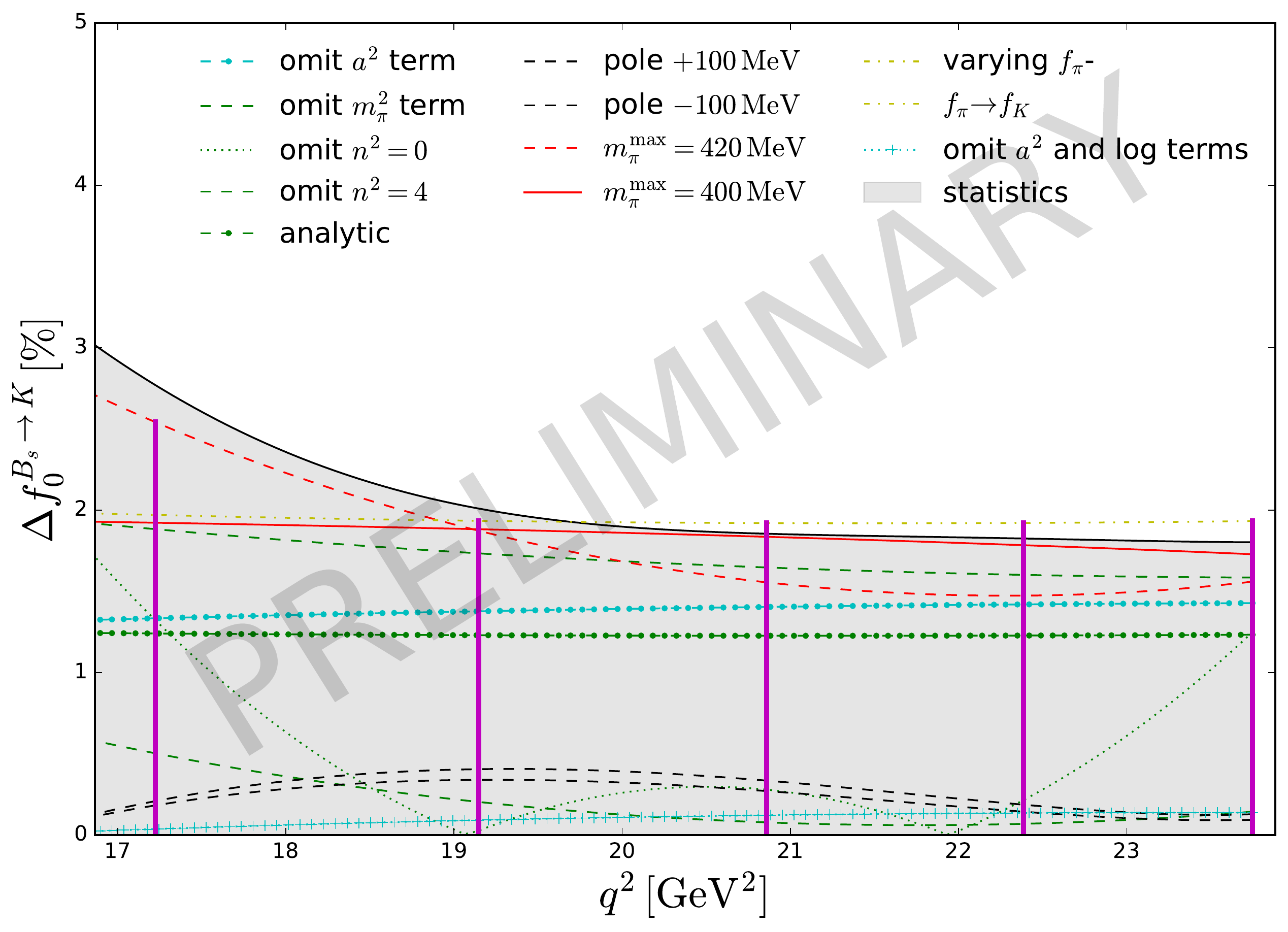}
  \caption{Preliminary estimate of systematic uncertainties for the determination of $B_s\to K \ell\nu$ form factors due to variations of the fit ansatz which is based on SU(2) heavy meson chiral perturbation theory. Vertical magenta lines indicate the $q^2$ values used in the subsequent $z$ expansion.}
  \label{fig.BsK_error}
\end{figure}

The $B_s\to K\ell \nu$ analysis follows the $B\to \pi\ell\nu$ analysis
presented in Subsection~\ref{Sec.Bpi}. Since only one light $u$ quark
enters in the valence sector, the chiral extrapolation is milder and
the data is statistically more precise as can be seen in
Fig.~\ref{fig.BsK_ChiPT}. As above, the colored symbols refer to the
form factors determined on our six ensembles and the colored lines
show the outcome of the chiral-continuum extrapolation at finite
lattice spacing and unphysical light quark masses whereas the black
line with gray error band shows the result in the chiral-continuum
limit. For this process we again use a fit ansatz based on SU(2) heavy
meson chiral perturbation theory and consider the hard kaon limit. The
fit function is obtained from Eq.~(\ref{Eq.HMXPT}) by substituting
$\pi$ with $K$ and using 
\begin{align}
\delta_f^{B_s\to K}=-\frac34 M_\pi^2 \log\left(\frac{M_\pi^2}{\Lambda^2}\right).
\end{align}
Further a pole factor is present for both form factors, $f_\perp$ and $f_\parallel$,
in the case of $B_s\to K\ell \nu$ extrapolations.
Fig.~\ref{fig.BsK_ChiPT} shows in addition, with black crosses, the synthetic
data points we will use subsequently to perform the kinematical extrapolation
over the full range of momentum transfer. As a first step to estimate systematic
uncertainties, we perform variations of our fit ansatz. In
Fig.~\ref{fig.BsK_error} we show relative deviations over our preferred fit as a
function of the $q^2$ range covered by our data. The gray error band indicates
the statistical uncertainty of our preferred fit and the colored lines indicate
different variations. Variations of our fit ansatz include adding higher order
or removing terms from the fit function, applying different
cuts to our data, or varying external parameters entering the fit function. As
is shown in the figure, most of these variations change the central value by
less than our statistical uncertainty. In addition we need to estimate
systematic effects originating, for example, from the lattice spacing, RHQ
parameters, strange quark mass, or renormalization coefficients.  Some of
these uncertainties are already in place, work is in progress to estimate
others. Here we use a preliminary estimate to obtain synthetic data points with
a combined statistical and systematic uncertainty. These data points are the
input for the $z$-expansion presented in Sec.~\ref{Sec.zfits_BsK}.

\subsection{$B_s\to D_s \ell \nu$} \label{Sec.BsDs}

\begin{figure}[tb]
  \includegraphics[width=0.48\textwidth]{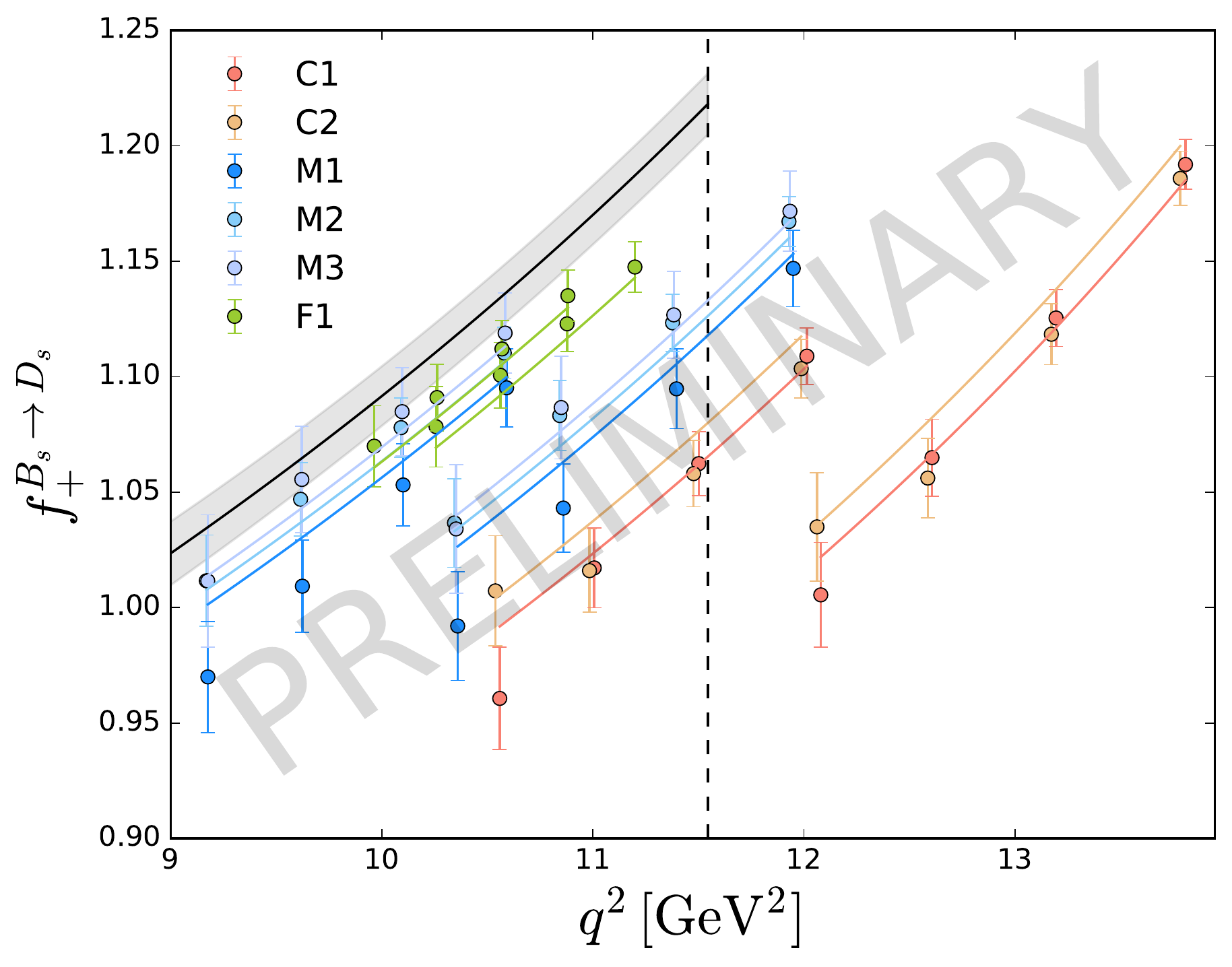}
  \includegraphics[width=0.48\textwidth]{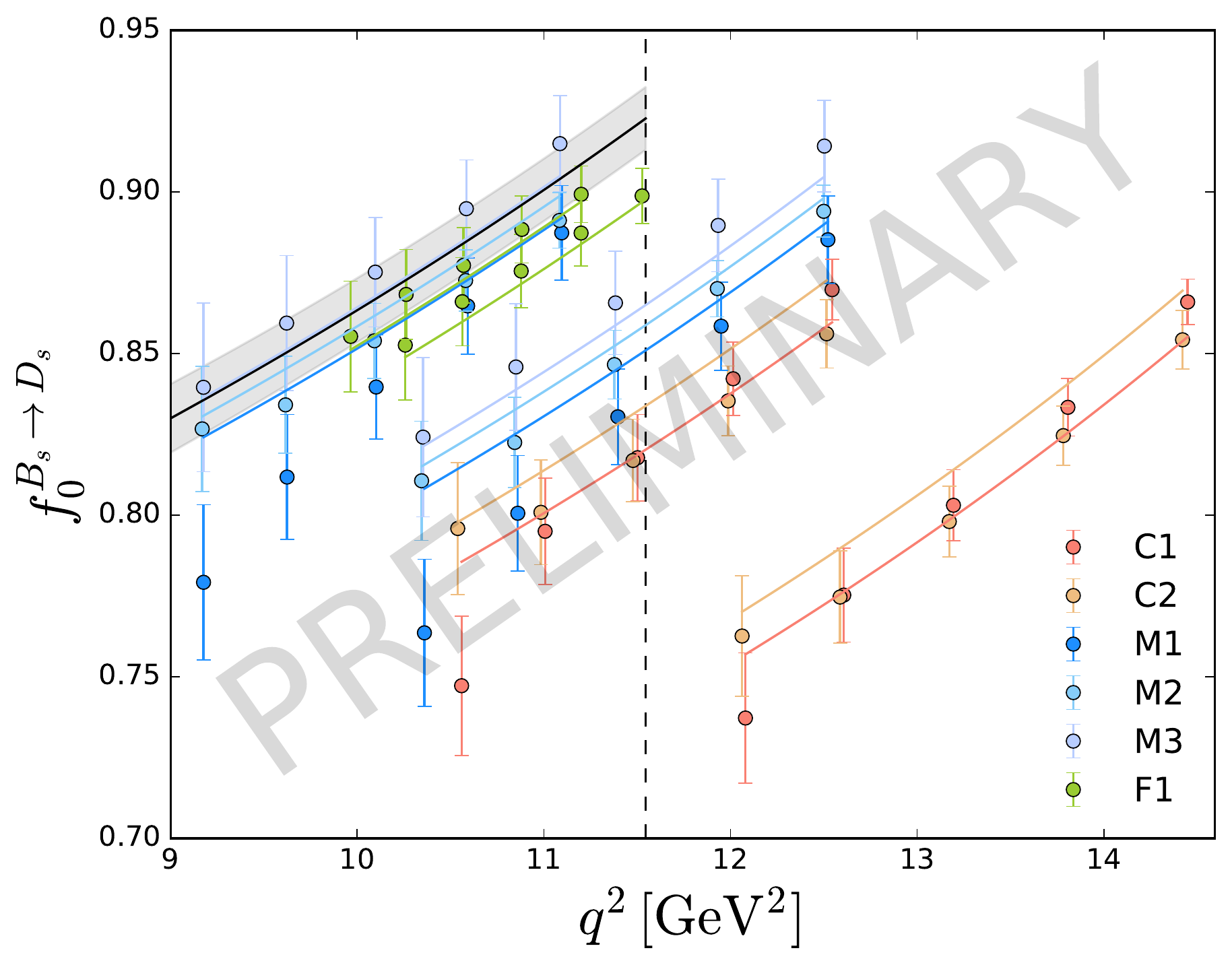}
  \caption{Global fit extrapolating our $B_s\to D_s \ell \nu$ form factors to physical quark masses and the continuum limit (gray band). The colored lines show the fit result at the different unphysical charm quark masses.}
  \label{fig.BsDs_ChiPT}
\end{figure}

\begin{figure}[tb]
  \includegraphics[width=0.48\textwidth]{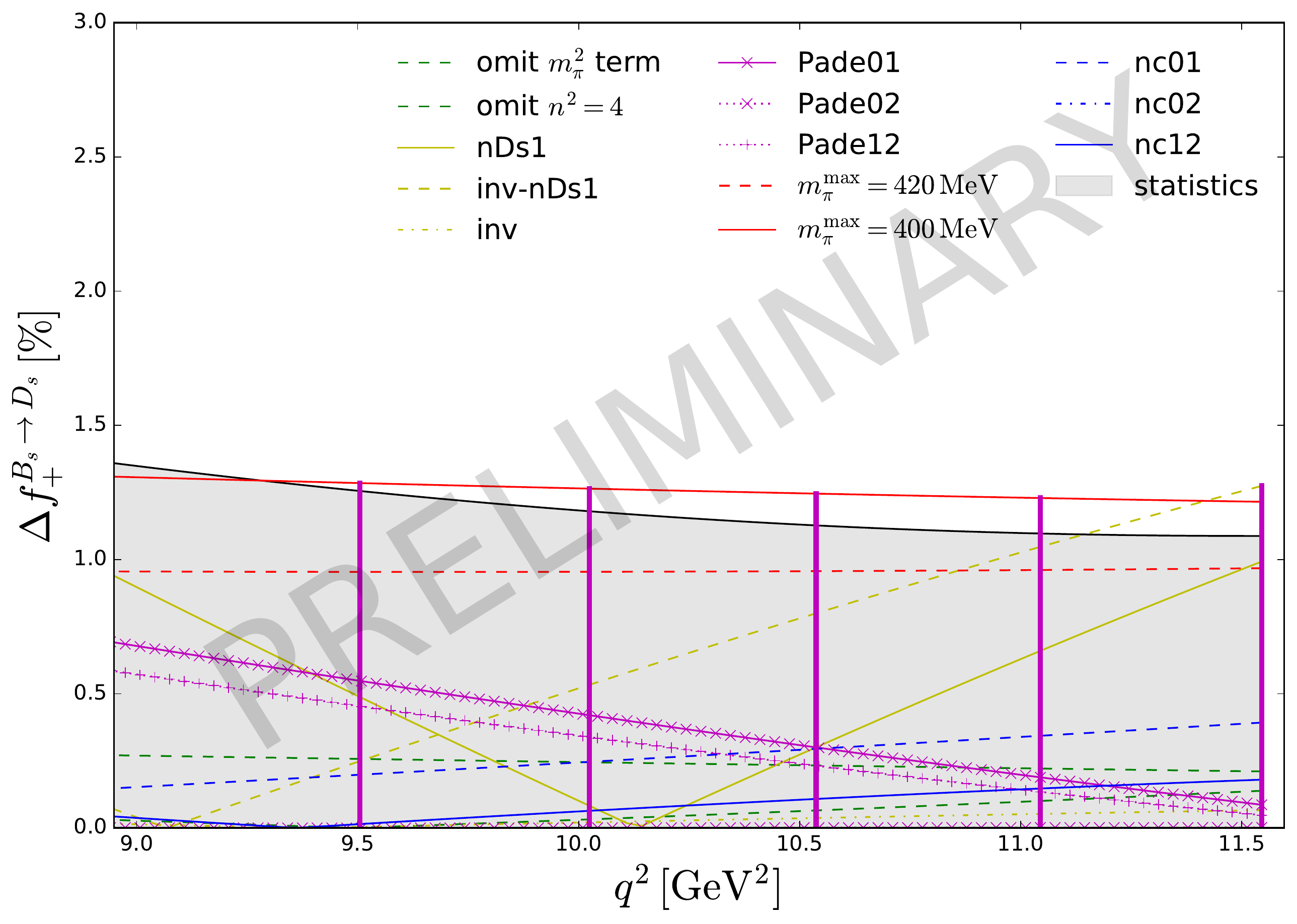}
  \includegraphics[width=0.48\textwidth]{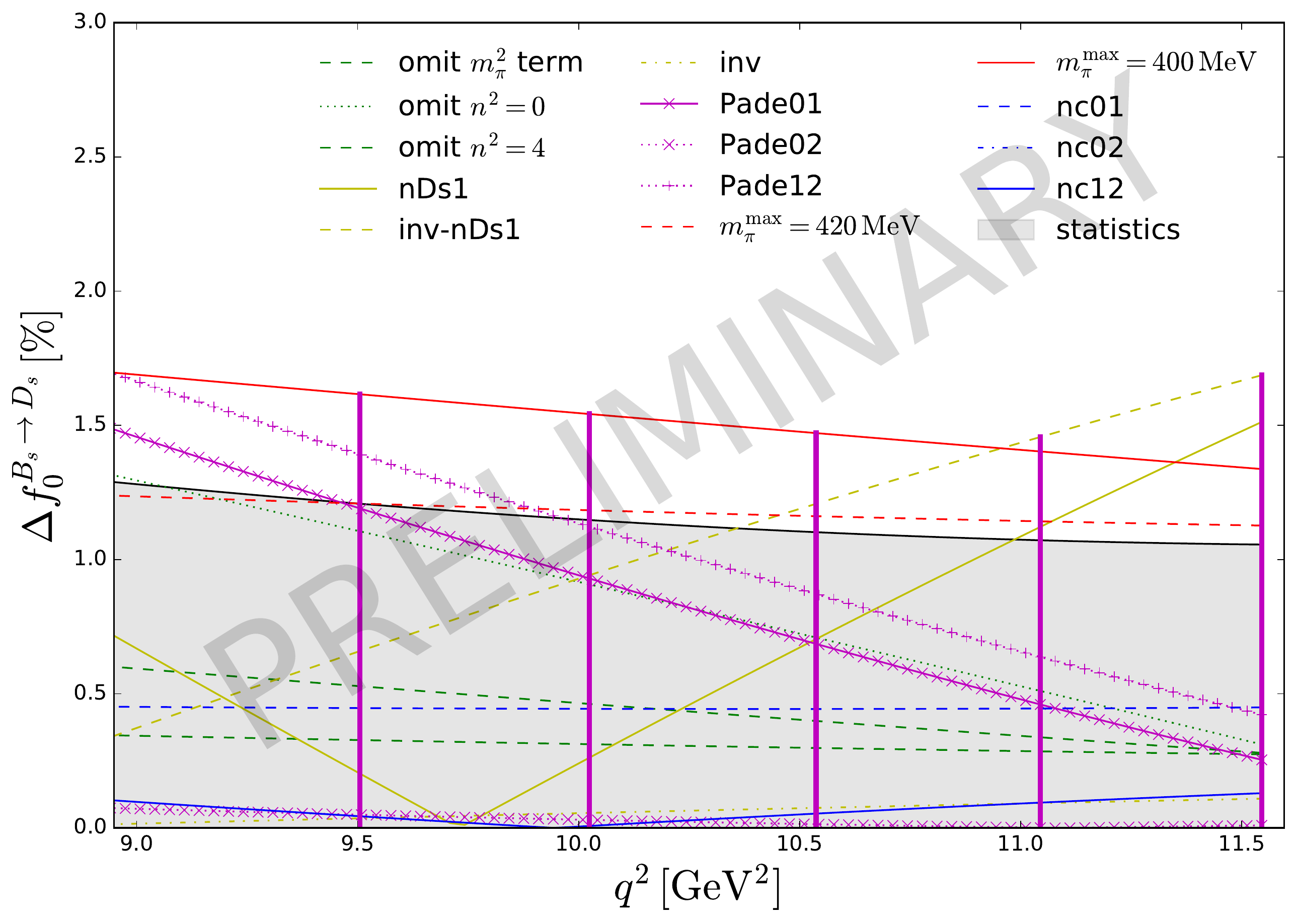}
  \caption{Preliminary estimate of systematic uncertainties for the determination of $B_s\to D_s \ell\nu$ form factors due to variations of the fit ansatz.  Vertical magenta lines indicate the $q^2$ values used in the subsequent $z$ expansion.}
  \label{fig.BsDs_error}
\end{figure}

For the $B_s\to D_s\ell\nu$ form factors, we simulate a set of charm
quark masses and then extra(inter)polate to the physical value, as
described in Subsection~\ref{Sec.BD}. Hence in
Fig.~\ref{fig.BsDs_ChiPT} we again show a set of data points for each
of our six ensembles. Since the $B_s\to D_s\ell\nu$ decays do not have
light quarks in the valence sector, we perform a global fit based on a
Pad\'e approximation in order to extrapolate to physical quark masses
and the continuum limit
\begin{gather}
        f(q^2,a, M_\pi, M_{D_s}) = \left[c_0 +c_1 \frac{M_\pi^2}{\Lambda^2} + \sum_{j=1}^{n_{D_s}}c_{2,j} \left[\Lambda \cdot \Delta M_{D_s}^{-1}\right]^j+ c_3 (a\Lambda)^2\right] P_{\alpha,\beta}(q^2/M_{B_s}^2), \\
       \text{with} \quad\Delta M_{D_s}^{-1} \equiv \bigg(\frac{1}{M_{D_s}} - \frac{1}{M_{D_s}^\text{phys}}\bigg)\quad\text{and}
        \quad  P_{\alpha.\beta}(x) = \frac{1+\sum_{i=1}^{N_\alpha} \alpha_i x^i}{1+\sum_{i=1}^{N_\beta} \beta_ix^i}. \nonumber
\end{gather}
$\Lambda$ is again a reference scale of of order $1\gev$.
The colored lines in Fig.~\ref{fig.BsDs_ChiPT} show the fit result at
unphysical charm quark mass and the gray band the form factors at
physical quark masses in the continuum. As for the $B_s\to K\ell\nu$
decays described above, we perform variations of our fit ansatz to
estimate its systematic uncertainty. Examples are shown in
Fig.~\ref{fig.BsDs_error}. Estimates of other systematic effects are
ongoing and we therefore use a preliminary error budget as input for
the $z$-expansion shown in Subsection~\ref{Sec.zfits_BsDs}.

\section{Kinematical $z$-expansion}
\label{Sec.zfits}

As shown in the previous Section, the lattice calculation of
semileptonic form factors for $B_{(s)}$ decays directly covers the
region of large momentum transfer (large $q^2$) and is most precise
near $q^2_\text{max}$. To extend the range to small momentum transfer,
we perform an additional, kinematical extrapolation typically referred
to as a $z$-expansion. As input for the $z$-expansion we use
`synthetic' data points extracted from our results after extrapolation
to physical quark masses and to the continuum. For these points we
estimate full systematic uncertainties and hence obtain a set of
form-factor data points at specific $q^2$ values with full,
statistical and systematic, uncertainties to be used for the
kinematical extrapolation. Although the systematic error budget is not
yet complete, we outline the next steps of the analysis.

In a first step the complex $q^2$ plane with a cut for $q^2\geq t_+$ is
 mapped onto the unit disk in $z$ with the transformation
\begin{align}
z(q^2, t_0) = \frac{\sqrt{1 - q^2/t_+} - \sqrt{1 - t_0/t_+}}{\sqrt{1 - q^2/t_+} + \sqrt{1 - t_0/t_+}}\,,
\end{align}
where $t_-= (M_{B_{(s)}}- M_P)^2$ for $P=\pi,\,K\,D,\,D_s$, and $t_+$ is
 fixed by the appropriate two-particle production threshold. In the case of 
$B\to\pi\ell\nu$ and $B_s\to K\ell\nu$, the start of the cut is given by
$t_+ = (M_B +M_\pi)^2$, whereas for $B\to D\ell\nu$ and $B_s\to D_s\ell\nu$ 
the cut begins at $t_+ = (M_B +M_D)^2$. Using $t_+$ and $t_-$ we define
$t_0 = t_+ - \sqrt{t_+(t_+-t_-)}$. Commonly two different implementations of the
$z$-expansion are considered for our processes of interest. Boyd, Grinstein, and
Lebed (BGL)~\cite{Boyd:1994tt} express the form factors ($X=+,0$) as
\begin{equation}
  \label{Eq.BGL}
f_X(q^2) = \frac{1}{B_X(q^2)\phi_X(q^2, t_0)}\sum_{n=0}^{N}a_n(t_0)z^n\,,
\end{equation}
where $\phi_X$ are outer functions and $B_X$ are the Blaschke factors
which vanish at the positions of sub-threshold poles (so that the
remaining $z$-dependence can be expanded in a power series). The $a_n$
coefficients are subject to a constraint derived from a dispersive
bound
\begin{equation}
  \sum_{n=0}^\infty a_n^2(t_0)\le 1.
  \label{Eq.ConstraintBGL}
\end{equation}  
An alternative implementation is given by Bourrely, Caprini, and
Lellouch (BCL)~\cite{Bourrely:2008za} who, for $X=+$ in $B\to\pi$
semileptonic decay, parametrized the form factor by
\begin{equation}
  f_X(q^2) = \frac{1}{1 - q^2/M^2_\text{pole}}\; \sum_{k=0}^{K} b_{k}(t_0)  z^k,
  \label{Eq.BCL}
\end{equation}
with the constraints
\begin{equation}
  \sum_{n=0}^\infty a_n^2 = \sum_{j=0}^K\sum_{k=0}^K b_j B_{jk} b_k \le 1
\qquad\text{and}\qquad
\sum_{k=0}^K(-1)^{k-1}kb_k(t_0)=0.
\label{Eq.ConstraintBCL}
\end{equation}
The first constraint in Eq.~(\ref{Eq.ConstraintBCL}) is derived by
equating the expressions in~(\ref{Eq.BGL}) and~(\ref{Eq.BCL}) and
expanding the known functions in powers of $z$ in order to relate the
$a_n$ and $b_n$ and hence determine the $B_{jk}$. The second
constraint applies only to $f_+$ and originates from angular momentum
conservation.


\subsection{$B_s\to K \ell\nu$}
\label{Sec.zfits_BsK}

\begin{figure}[tb]
  \includegraphics[width=0.48\textwidth]{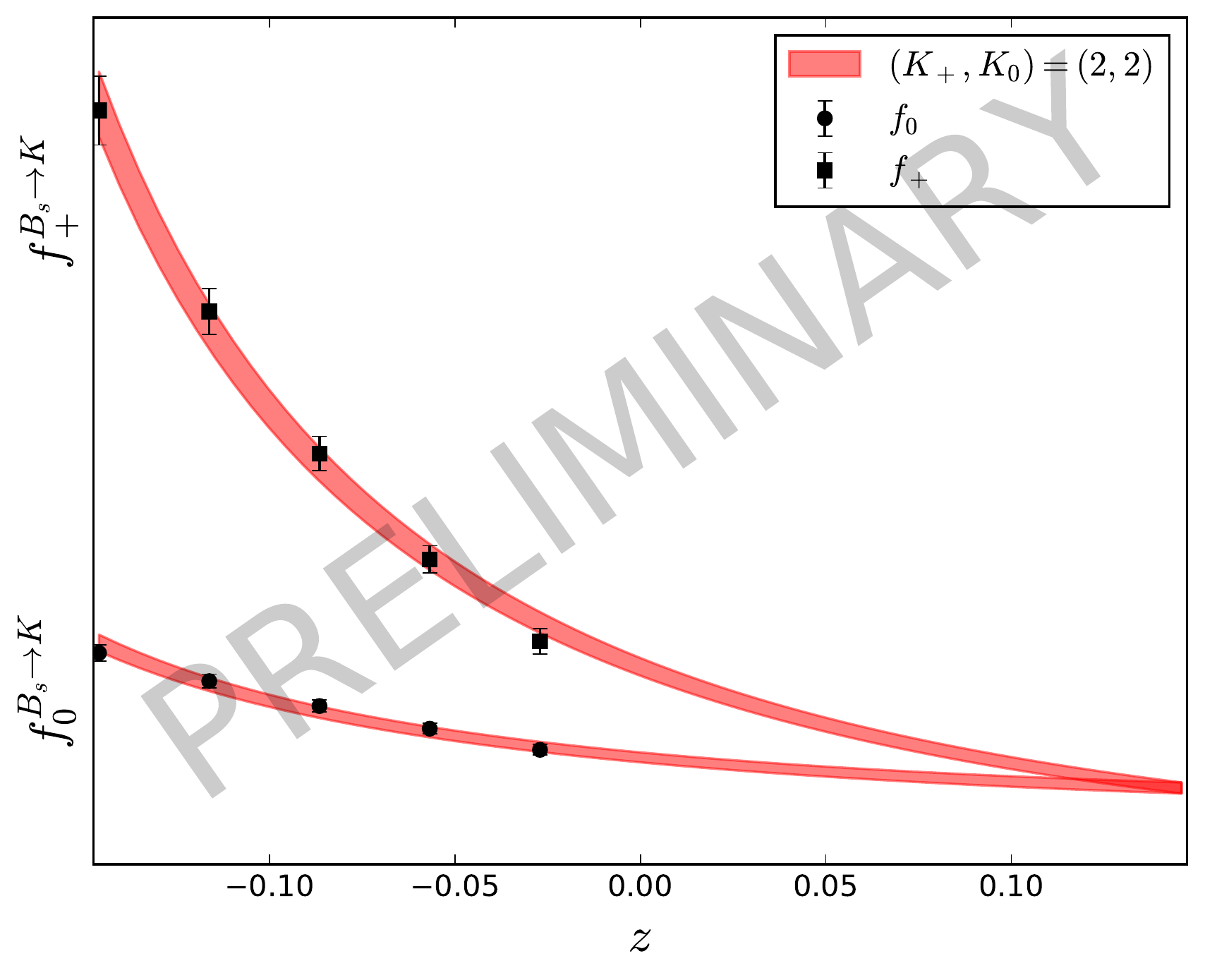}
  \includegraphics[width=0.48\textwidth]{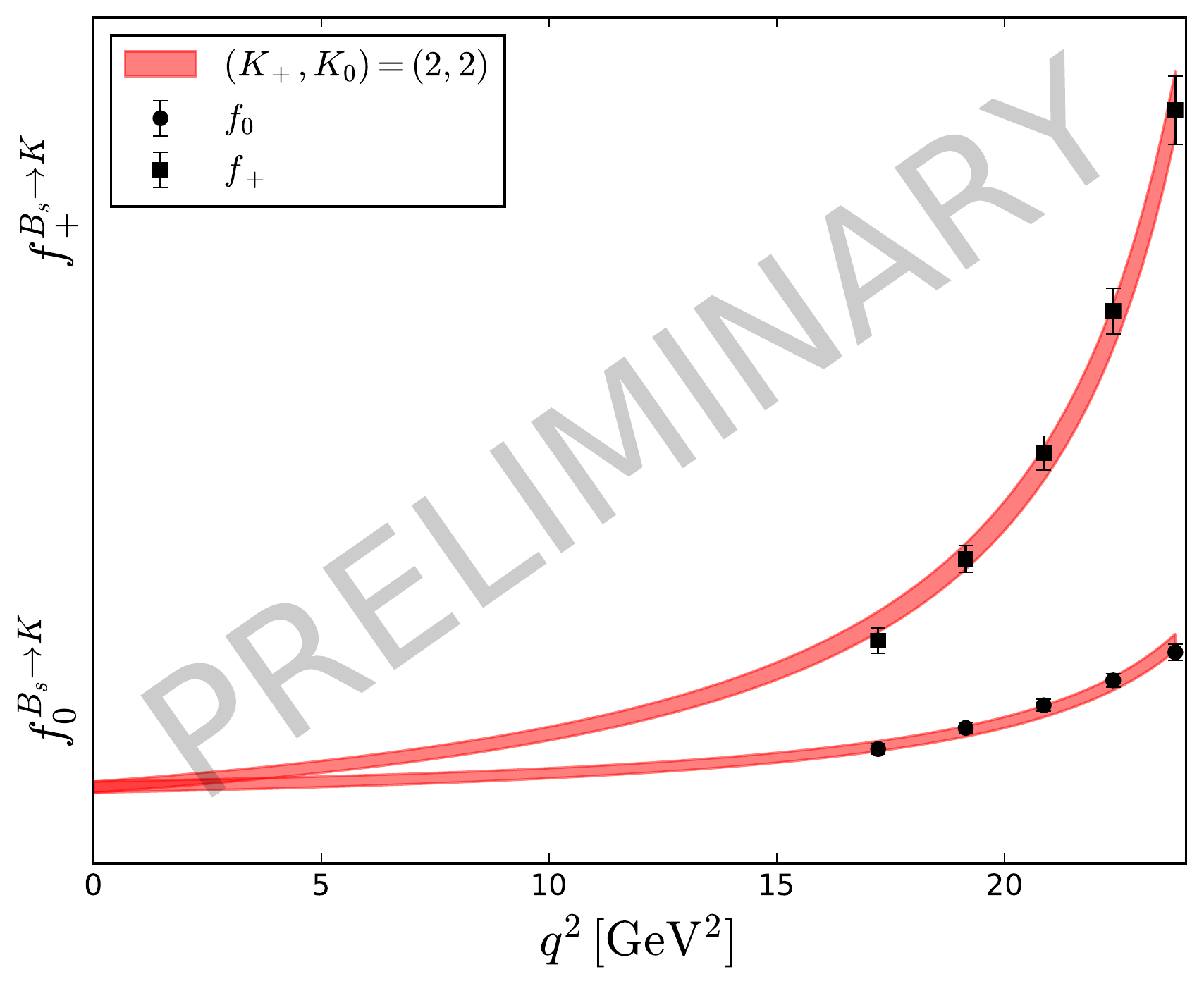}
  \caption{Preliminary kinematical extrapolation ($z$-expansion) of our $B_s\to K \ell\nu$ form factors over the full $q^2$ range using the BCL parametrization. The extrapolation is performed using a set of five synthetic data points (black symbols) which are obtained with a non-final error budget.}
  \label{fig.BsK_zfit}
\end{figure}

Using our results of Section \ref{Sec.BsK} with a preliminary error
budget for the synthetic data points, we perform a kinematical
extrapolation down to $q^2=0$. We use the BCL parametrization 
described above with a pole mass $M_\text{pole} = M_{B^*} = 5.33\gev$ 
for $f_+$, and no pole for $f_0$, since the theoretically predicted 
$B^*(0^+)$ mass, $5.63\gev$~\cite{Bardeen:2003kt} is above $M_B+M_\pi$. The
outcome is shown in Fig.~\ref{fig.BsK_zfit}. On the left we present
the resulting form factors in $z$-space, on the right we convert back
to $q^2$.
 
Once our systematic errors have been finalized, we have everything in
place to obtain the form factors with full $q^2$ dependence.
Integrating these form factors over $q^2$, we can derive predictions
to test the universality of lepton flavors
\begin{align}
  R_{K}^{\tau/\mu}\equiv \frac{BF(B_s\to K\tau\nu_\tau)}{BF(B_s\to K \mu \nu_\mu)}.
\end{align}
or compare our results to the determination by HPQCD \cite{Bouchard:2014ypa}, Alpha \cite{Bahr:2016ayy}, or Fermilab/MILC \cite{Bazavov:2019aom}.


\subsection{$B_s\to D_s \ell\nu$}
\label{Sec.zfits_BsDs}

\begin{figure}[tb]
  \includegraphics[width=0.48\textwidth]{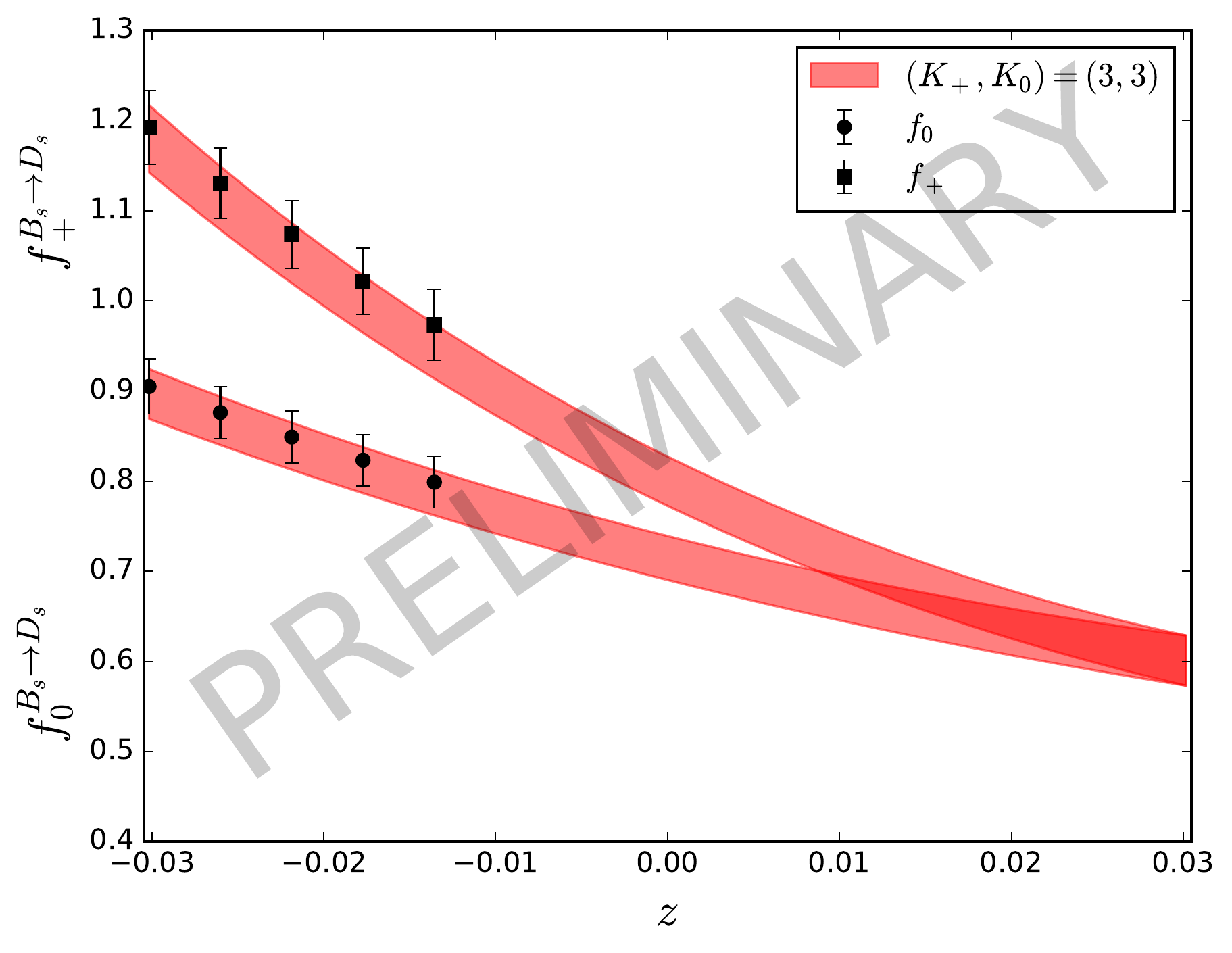}
  \includegraphics[width=0.48\textwidth]{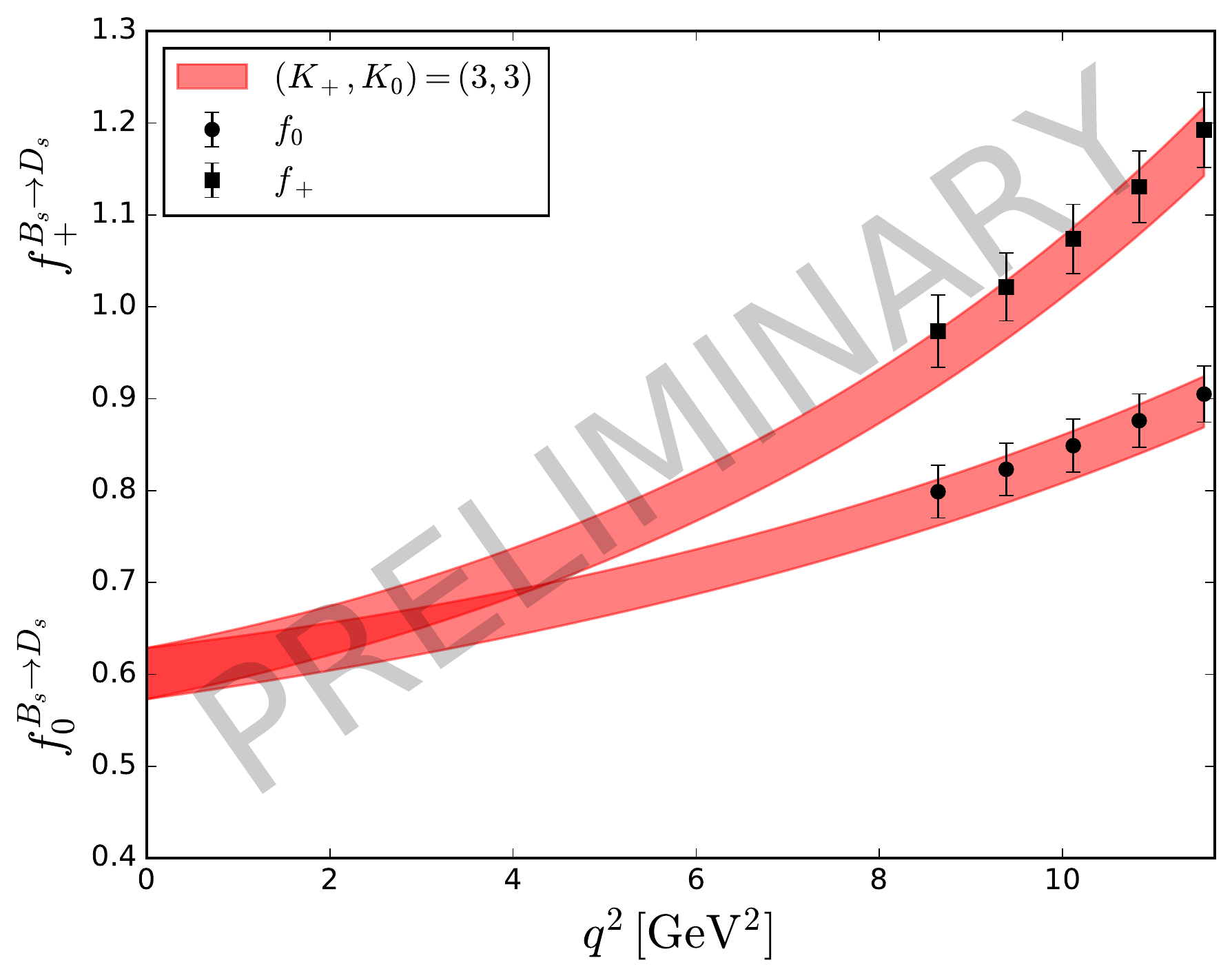}
  \caption{Preliminary kinematical extrapolation ($z$-expansion) of our $B_s\to D_s \ell\nu$ form factors over the full $q^2$ range using the BCL parametrization. The extrapolation is performed using a set of five synthetic data points (black symbols) which are obtained with a non-final error budget.}
  \label{fig.BsDs_zfit}
\end{figure}

In the case of $B_s\to D_s\ell\nu$ decays we proceed similarly. Starting from the synthetic data points with preliminary error budget obtained in Section \ref{Sec.BsDs}, we perform a BCL-style $z$-expansion using at present pole masses $M_+= M_{B_c^*} =6.33$ GeV and $M_0 = 6.69$ GeV \cite{Eichten:2019gig}.  Figure \ref{fig.BsDs_zfit} shows again on the left the resulting form factors vs.~$z$ and on the right vs.~$q^2$.  A finalized error budget will allow us to compare our results to the determination by HPQCD \cite{Monahan:2017uby,McLean:2019qcx} and Fermilab/MILC \cite{Bailey:2012rr,Bazavov:2019aom}. In addition we can calculate 
\begin{align}
  R_{D_s}^{\tau/\mu}\equiv \frac{BF(B_s\to D_s\tau\nu_\tau)}{BF(B_s\to D_s \mu \nu_\mu)},
\end{align}
to test lepton flavor universality.

\section{Summary}
We have reported updates on the RBC-UKQCD $B$-physics program which currently has the main focus to calculate form factors of semileptonic decays. Our set-up features $b$-quarks simulated with the RHQ action but uses DWF for up/down, strange, and charm quarks. Here we presented the status of our analysis for $B\to\pi\ell\nu$, $B\to D\ell\nu$, $B_s\to K\ell\nu$, and $B_s\to D_s\ell\nu$. In addition to the charged current decays with pseudoscalar final state, our framework also includes operators to determine $W^\pm$ mediated decays to vector final states as well as Glashow-Iliopoulos-Maiani  (GIM) suppressed neutral current decays \cite{Flynn:2015ynk,Flynn:2016vej}.

In parallel we are continuing our efforts to advance the use of heavy DWF for $b$ quarks and extend the methods used in the calculation of decay constants \cite{Boyle:2017jwu} or neutral meson mixing matrix elements \cite{Boyle:2018knm} to semileptonic decays \cite{Boyle:2019cdl}.

\section*{Acknowledgments}
We thank our RBC and UKQCD collaborators for helpful discussions and
suggestions, and P.~Gambino for discussion on unitarity-constrained
fits. Computations were performed on resources provided by the USQCD
Collaboration, funded by the Office of Science of the U.S.~Department
of Energy, on the ARCHER UK National Supercomputing Service
(\href{http://www.archer.ac.uk}{http://www.archer.ac.uk}), as well as
on computers at Columbia University and Brookhaven National
Laboratory. We used gauge field configurations generated on the DiRAC
Blue Gene~Q system at the University of Edinburgh, part of the DiRAC
Facility, funded by BIS National E-infrastructure grant ST/K000411/1
and STFC grants ST/H008845/1, ST/K005804/1 and ST/K005790/1. This
project has received funding from STFC grants ST/L000458/1 and ST/P000711/1. RH was
supported by the DISCnet Centre for Doctoral Training (STFC grant
ST/P006760/1). AS was supported in part by US DOE contract DE--SC0012704. OW acknowledges support from DOE grant DE--SC0010005. No new experimental data was generated.
{\small
\bibliography{../B_meson}
\bibliographystyle{JHEP-notitle}
}

\end{document}